# Hybrid Power-Line/Wireless Communication Systems For Indoor Applications

Waled Gheth[1], Khaled M. Rabie[1], Bamidele Adebisi[1], Muhammad Ijaz[1], Georgina Harris[1] and A. Alfitouri[2]
[1]School of Engineering, Manchester Metropolitan University, UK
[2]School of Electrical and Electronic Engineering, University of Manchester, UK
Emails:{w.gheth, k.rabie, b.adebisi, m.ijaz, g.harris}@mmu.ac.uk; abdurrahman.alfitouri@manchester.ac.uk

*Abstract*—The high attenuation which increases as frequency increases in power line communication systems (PLC) makes achieving and sustaining high data rates over such channels a challenging task. In most communication systems, relays are normally deployed to overcome the effect of signal attenuation and to achieve reliable transmissions between users. Amplify-and-forward (AF) and decode-and-forward (DF) relays are often used. In contrast to the existing work on this topic which advocate utilizing relays either between PLC links or between wireless links, the authors of this paper focus on using AF relaying for in-home PLC and wireless systems where one PLC user wants to share data with one wireless user which has no access to the PLC network. We refer to this system as the Hybrid approach. A mathematical method is developed for this network to formulate the capacity of the system by exploiting the statistical properties of both PLC and wireless channels. Monte Carlo simulations are used throughout this paper to validate the analytical results. The impact of several system parameters is investigated in this paper. For the sake of comparison, we also provide the performance analysis and results for a PLC only network. The results showed that the performance of the hybrid system enhances as we increase the relay gain and deteriorates as the end-to-end distance increases. It is also found that the PLC only scheme may outperform the hybrid one if the relay gain is relatively small; however, the user mobility offered by the hybrid system remains the main advantage over the PLC only system.

## I. INTRODUCTION

THE advantage of using the pre-installed infrastructure of wiring networks makes power-line communications (PLCs) a competitive technology for broad-band communications. Due to the horrible channel of this technology, there are several practical issues ought to be overcome in order to ensure reliable transmission through the PLC system. Electromagnetic interference to the surrounding wireless systems is one of the practical issues that needs to be considered when designing PLC systems [1]. Certain limits are specified by standardization bodies for the maximum allowable power that a PLC modem can transmit to make sure that surrounding wireless systems are not affected by the PLC system [1]. This might lead to low values of signal-to-noise (SNR) at the destination.

Furthermore, transmission over PLC channels is mostly affected by noise. Background noise and impulsive noise are the two main categories of noise over PLC systems. While the former has negligible variation over time and low power spectral density (PSD), the latter has a random occurrence, very short duration and high power spectral density (PSD) [2]. Moreover, the desirable date rate can significantly be degraded by the attenuation caused by the frequency-selectivity of the PLC channel which introduces inter-symbol interference (ISI). This attenuation is significant at very high frequencies and increases with increasing the transmitter–receiver distance [3], [4].

In the wireless domain, communication signals suffer from attenuation caused by the reflected waves. The receiver receives signals containing a huge number of the reflected waves among, which is known as multi path reception and this is how the indoor channel is characterized [5]. Attenuation and multi-fading can rapidly degrade the SNR or signal-to-interference ratio (SINR) [5]. Recently, integrating PLC systems with other communication systems, such as visible light communication (VLC) and wireless communication systems, has received considered attention from researchers [5]–[7]. The implementation of such integration would introduce several benefits such as capacity, security and high data rate increased [5], [7]. A cascaded PLC and wireless channels was analysed in [8]. The results showed that lower attenuation was offered in the PLC to wireless direction than the hybrid channel in the opposite direction. It was also found that PSD of the PLC channel is defferent from that in the wireless one.

Inspired by adopting other communication technologies for PLC systems, many researchers have studied the implementation of relays for PLC systems to improve the capacity and to tackle the signal attenuation issues [9], [10]. Amplify-and-forward (AF) and decode-and-forward (DF) are the two common types of relaying systems. While the former amplifies the received signal then re-transmits it to its destination, the latter decodes and re-encodes the received signal before forwarding it to the destination [11]. Other types of relays investigated over PLC channels include incremental DF (IDF) and selective DF (SDF) [12], [13].

The use of half duplex time division relaying protocols in in-home PLC systems was investigated in [10] and the results were promising showing a significant improvement consequently in data rate and power saving depending on the position of the relay in the network and the network size. Although the authors in [14] concluded that the capacity of indoor PLC systems can be considerably improved by positioning the relays in proper locations in the network, they reported that relaying could be more effective for wireless than it is for PLC networks. Authors in [9] reported that two-way relaying could provide more performance improvement than the one-way relaying would in in-home PLC systems. The implementation of DF relaying protocol for wireless-powerline hybrid communication system was investigated in [15] and it was found that the considered system is significantly affected by the noise of the PLC link.

Implementation of multi-way relaying in parallel for indoor PLC systems was discussed in [16] and the achievable data rates were high compared to those achieved by direct transmission. The implementation of relays in hybrid communication systems in order to improve performance was also studied in [5] where the source transmits signals through PLC and wireless networks simultaneously to a destination through an AF relay. It was found that the use of hybrid systems outperforms the use of PLC or wireless systems individually.

In this framework we study the deploying relays to connect PLC with wireless link as no work in the literature has studied such deployment. Therefore, the authors of this work study the implementation of AF relaying to connect a cascade PLC and wireless links. The information signal is firstly transmitted by the PLC user to the relay through the PLC link, which is then amplified and forwarded by the AF relay via wireless channel to one of the wireless destinations. Analytical expressions for the total capacity of the system is derived in this paper taking into consideration the different characteristics of each link. Formulating the overall capacity of the hybrid PLC and wireless system offers the opportunities to study the performance of such cascade system and to examine the effect of the different system parameters such as relay location and relay-to-destination distance. Monte Carlo simulations were used throughout the paper to validate our analytical results. For comparison's sake, the performance of the PLC only system also analyzed. Results show that after a certain distance between the source and destination the proposed system can provide better performance compared to the PLC only system when the AF relay gain is sufficiently large.

The rest of the paper is organized as follows. The system model is described in Sec. II. The system performance analysis is presented in Sec. III. The discussion of the numerical results is in Sec. IV. Sec. V draws the conclusion of this paper.

## II. System Model

In this section, the system model for the hybrid PLC/wireless system with the use of relay is introduced and we formulate the overall capacity. The system model under consideration is shown in Fig.1. The practical applications of the proposed system could be at any building company, home or university where the source has access to the PLC link and the destination does not. PLC users can be connected to the network through PLC modems which exploit the power lines for communications. We assume that one PLC user is connected to one of the outlets and this user wants to share data with one of the wireless users in another room which has no access to the PLC network. A relay node is used in order to assist the data transmission. Data transmitted through a PLC channel is amplified and redirected to its receiver through a wireless link by an AF relay. The source-to-relay channel, i.e., the PLC link, is represented by the complex channel gain $h_P$ and $h_W$ denotes the complex channel gain for the wireless link (i.e., the relay-to-destination channel). Both channels are assumed to be independent and identically distributed with log-normal distribution and Rayleigh fading distribution for PLC and wireless channels, respectively [11], [17]. Because of the nature of the network structure adopted here, it is logical to assume that there is no direct link between the source and destination nodes. It is worth mentioning, for simplicity and without loss of generality, that noise over the two links is assumed to be additive white Gaussian noise (AWGN).

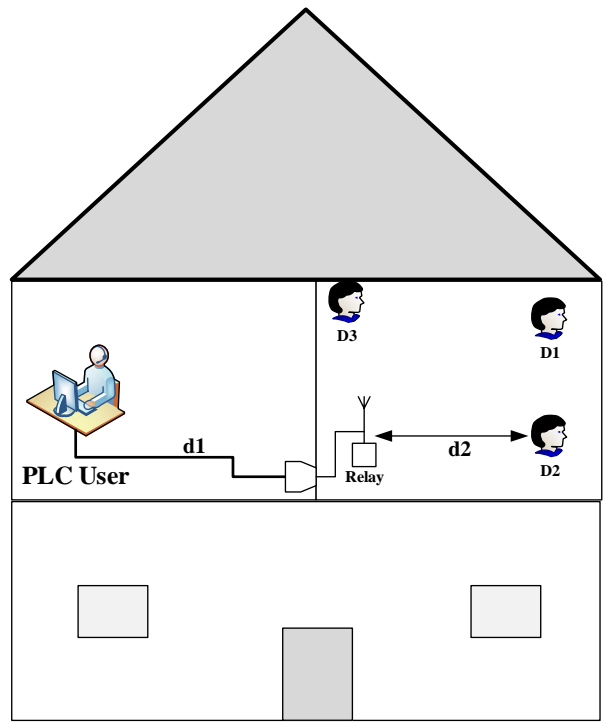

Fig. 1: System model for the hybrid PLC/Wireless network for indoor communication applications.

*a) Hybrid System Overall Capacity:* As we mentioned above, the overall communication is divided into two stages. The received signal by the relay from the PLC user during phase one can given by the following

$$y_r(t) = \sqrt{P_s} e^{-\alpha d_1} h_P s + n_r, \quad (1)$$

where $P_s$ is the source transmit power, $d_1$ is the PLC link length which represents the total distance from the PLC user to the relay, $s$ is the useful information with E[s]=1, $n_r$ is the noise at the AF relay which is considered to be complex Gaussian [18], [19] with variance $\sigma_r^2$ and zero mean, and $\alpha$ is the PLC channel attenuation factor and is given by $\alpha = a_0 + a_1 f^k$, where constants $a_0$ and $a_1$ are normally determined by measurements, $f$ is the system operating frequency and $k$ presents the the attenuation factor exponent.

In the second phase the following formula describes destination received signal

$$y_d(t) = h_w G \sqrt{P_r} d_2^{-\frac{m}{2}} \left( \sqrt{P_s} e^{-\alpha d_1} h_P s + n_r \right) + n_d, \quad (2)$$

where $G$ is the relay gain, $d_2$ is the relay-to-destination distance, $m$ denotes the path loss exponent of the wireless channel, $P_r$ is the relay transmit power and $n_d$ represents the destination noise with zero mean and variance $\sigma_d^2$.

Using (2), and grouping the information signal and noise signal terms, the destination can be written as

$$\gamma = \frac{G^2 P_r d_2^{-m} P_s e^{-2\alpha d_1} |h_P|^2 |h_w|^2}{G^2 P_r d_2^{-m} |h_w|^2 \sigma_r^2 + \sigma_d^2}. \quad (3)$$

The instantaneous data rate of a communication system can be calculated as

$$C = \frac{1}{2} E\left[\log(1+\gamma)\right], \quad (4)$$

while $\gamma$ is the SNR of the system and the factor $\frac{1}{2}$ is due to the fact that the whole communication process between the transmitter and receiver is functioned in two stages [20].

### III. Performance Analysis

In this section, we derive accurate analytical expressions for both the hybrid and PLC only systems.

#### A. PLC-Wireless System

To begin with, equation (3) can be rewritten as

$$\gamma = \frac{P_s e^{-2\alpha d_1} |h_P|^2}{\sigma_r^2 + \frac{\sigma_d^2}{P_r G^2 d_2^{-m} |h_w|^2}}. \quad (5)$$

To simplify our analysis, let us assume that $K = P_s e^{-\alpha d_1} |h_P|^2$, $L = \sigma_r^2$ and $M = \frac{\sigma_d^2}{P_r G^2 d_2^{-m} |h_w|^2}$. Now we can write (5) as

$$\gamma = \frac{K}{L+M}. \quad (6)$$

Substituting (6) into (4), we can be express the capacity of the hybrid system as

$$C_H = \frac{1}{2} E\left[\log\left(1 + \frac{K}{L+M}\right)\right], \quad (7)$$

where the superscript H denotes the hybrid system. According to [21, Eq (5)], for any non-negative random variable $u$ and $v$, we can calculate the average capacity as follows

$$E\left[\ln(1+\frac{u}{v})\right] = \int_0^\infty \frac{1}{z}\left(1 - M_u(z)\right)\mathcal{M}_v(z)dz. \quad (8)$$

Using this definition, we can rewrite (7) as follows

$$E\left[\ln(1+\frac{K}{L+M})\right] = \frac{1}{2}\int_0^\infty \frac{1}{z}\left(1 - M_K(z)\right)M_{L+M}(z)dz, \quad (9)$$

where $\mathcal{M}_K(z)$ and $\mathcal{M}_{L+M}(z)$ are the moment generation functions (MGFs) of the random variables $K$ and $L+M$, respectively.

$$\mathcal{M}_K(z) = M_{|h_P|^2} P_s e^{-2\alpha d_1}. \quad (10)$$

As mentioned before that the PLC channel is represented by log-normal distribution. According to [22, Eq (2.54)] the MGF OF the log-normal distribution is given by

$$\mathcal{M}_h(z) \simeq \frac{1}{\sqrt{\pi}} \sum_{n=1}^{N_p} H_{x_n} \exp\left(10^{\frac{-\sqrt{2}\sigma_h x_n + (-2\mu_h)}{10}} z\right), \quad (11)$$

where $H_{x_n}$ represents the weight factors of the $N_p$ order Hermite polynomial and $x_n$ is the zeros of the same order. By using the previous equation for $|h_P|^2$ with log-normal parameters $(2\mu_{h_P}, 4\sigma_{h_P}^2)$, the MGF of $K$ can be expressed as

$$\mathcal{M}_K(z) \simeq \frac{1}{\sqrt{\pi}} \sum_{n=1}^{N_p} H_{x_n} \exp\left(10^{\frac{-\sqrt{2}2\sigma_{h_P} x_n + (-2\mu_{h_P})}{10}} P_s e^{-2\alpha d_1} z\right). \quad (12)$$

On the other hand, the wireless channel is considered to be affected by independent and identical distributed complex Gaussian fading which has zero mean and variance equal to 1. Therefore, $|h_w|$ follows Rayleigh distribution and $|h_w|^2$ becomes exponentially distributed. As a result, the MGF of $L+M$ can be expressed as

$$\mathcal{M}_{L+M}(z) = e^{z\sigma_r^2} 2\sqrt{\frac{z\sigma_d^2}{P_r G^2 d_2^{-m}}} I_K\left(1, 2\sqrt{\frac{z\sigma_d^2}{P_r G^2 d_2^{-m}}}\right), \quad (13)$$

where $I_K$ is the modified Bessel function of the first kind.

#### B. PLC Only System

In this subsection, we assume that the transmitter and the receiver are only connected through a PLC link with no relay in between. Therefore, the signal at the destination can be given by

$$y_d(t) = \sqrt{P_s} e^{-\alpha d} h s + n_d, \quad (14)$$

where $d$ is the PLC link length which represents the total distance of the end-to-end nodes and $n_d$ is the destination noise which is assumed to be complex Gaussian [18], [19] with zero mean and variance $\sigma_d^2$,

From (17), the SNR at the destination can be written as follows

$$\gamma = \frac{P_s e^{-2\alpha d} |h|^2}{\sigma_d^2}. \quad (15)$$

Similar to the case with the hybrid system, the overall capacity is given by

$$C = \frac{1}{2} E\left[\log(1+\gamma)\right]. \quad (16)$$

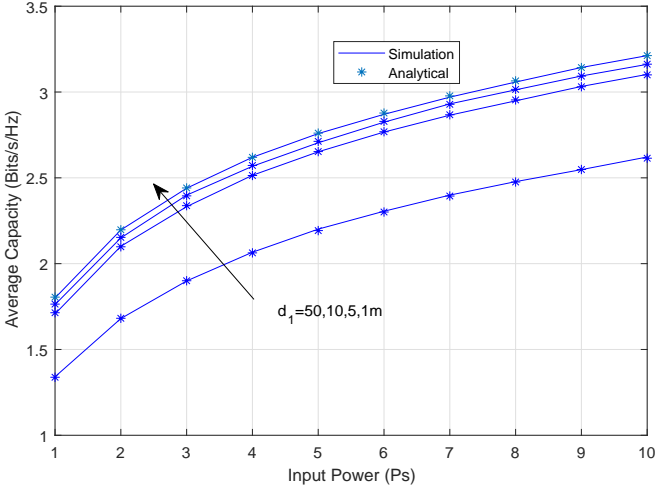

Fig. 2: Average capacity versus input power for different values of source-relay distance.

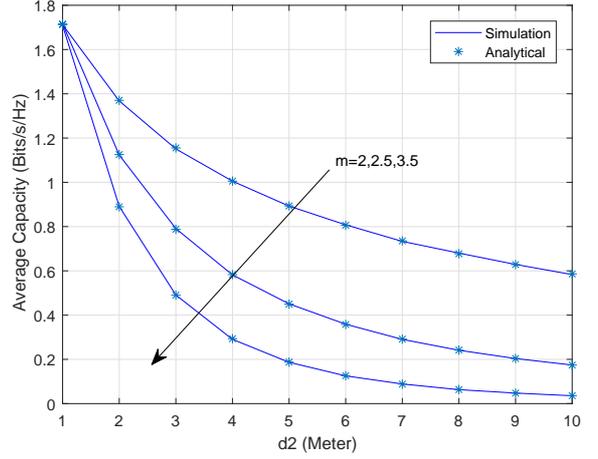

Fig. 3: Average capacity versus relay-destination distance for different values of path loss exponent of the wireless channel.

Mathematically, we can rewrite (16) as follows

$$C = \int_0^\infty \log_2(1+\gamma) f_\gamma(\gamma) d\gamma, \quad (17)$$

where $f_\gamma(\gamma)$ is the probability density function (PDF) of $\gamma$. By applying the PDF formula of the log-normal distribution, PDF can be expressed as

$$f_\gamma(\gamma) = \frac{\zeta}{z\sqrt{8\pi}\sigma} \exp\left(-\frac{(\zeta\ln(\gamma) - (2\mu + \zeta\ln(a_1)))^2}{8\sigma^2}\right), \quad (18)$$

where $\zeta$ is the scaling constant and it is equal to $10/\ln(10)$ and $a_1 = \frac{P_s e^{-2\alpha d}}{\sigma_{d2}^2}$. Hermite–Gauss quadrature is used in order to get an accurate approximation for (17). Assuming that

$$x = \frac{\zeta\ln(\gamma) - 2\mu + \zeta\ln(a_m)}{8\sigma^2}. \quad (19)$$

Now, we can rewrite (17) as

$$E[C_P] = \int_\infty^\infty \frac{1}{\pi} h(x) \exp[-x^2] dx. \quad (20)$$

By using Hermite–Gauss quadrature (20) can be expressed as

$$C_P = \sum_{n=1}^{N_p} \frac{1}{\sqrt{\pi}} H_{x_n} h(x_n), \quad (21)$$

where

$$h(x_n) = \log_2\left(1 + \exp\left(\frac{\sqrt{8}\sigma x_n + 2\mu + \zeta ln(a_1)}{\zeta}\right)\right). \quad (22)$$

It is worth noting that similar analysis appeared in [23].

## IV. NUMERICAL RESULTS

Monte Carlo Simulations are used in this section to verify the obtained numerical results for the overall capacity of the system. The considered system parameters are, unless indicated otherwise, as follows: operating frequency of the system $f = 500$KHz, exponent of the attenuation factor of the PLC link $k = 0.7$, $a_0 = 2.03 \times 10^{-3}$, $a_1 = 3.75 \times 10^{-7}$, $d_1 = 10$m, the input power $P_s = 1$W, $P_r = 1$W, relay gain $G = 1$, relay to destination distance $d_2 = 1$m and input SNR is 10dB [11], [24].

To explore the effect of the distance between the source and the relay on the system performance, the overall capacity is plotted in Fig. 2 versus the input power for different values of PLC link lengths. As we can see that there is a perfect agreement between the results obtained by Monte Carlo Simulation and the analytical results, obtained from 9. It is also clear from the figure the capacity improves as the input power increases and the source-relay distance decreases. Generally speaking, the average capacity of the system is almost the same when the source-to- relay distance changes from 1 to 10m particularly for small values of the input power. On the other hand, a performance degradation can be clearly seen when the distance between the source and relay is increased to 50m.

In Fig. 3, the total capacity is plotted against the distance between the relay and the destination with different values of $m$. As anticipated, we can see that the average capacity is degraded as the relay-destination distance and $m$ increase. In general, when $m = 2$, which is the minimum value of the path loss exponent considered in the figure the system has the best performance compared to the other values of $m$. The system has the worst performance when $m = 3.5$ which represents the highest value of the path loss exponent considered in the figure. It is also noticeable that the performance of the system is high for shorter distance between the relay and destination compared to the performance of the system for longer distance.

The behavior of the total capacity when the gain of the relay increases is illustrated in Fig. 4. It is obvious that the performance of the system enhances significantly as the relay gain increases and this justifies the main purpose of the relay use in the system. As seen in the figure, the average capacity is only 0.9Bits/s/Hz when the gain of the relay is 1dB and the relay-destination distance is 5 m, while it becomes almost 2.2 Bits/s/Hz when the gain increases to 10dB for the same relay-destination distance. On the other hand, the system becomes more efficient when the distance between the relay and the destination is smaller.

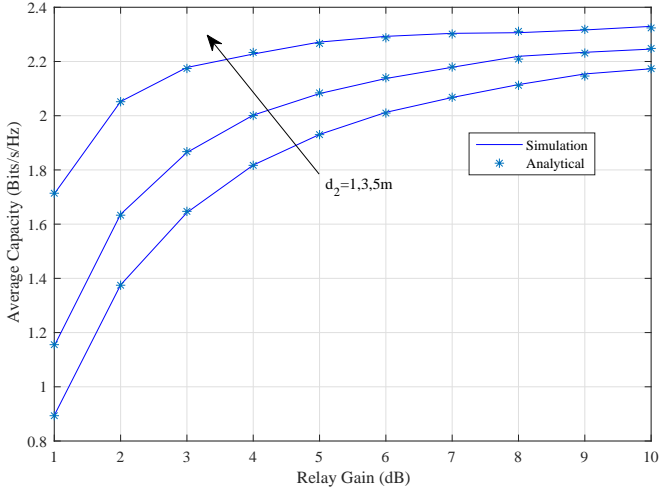

Fig. 4: Average capacity versus the relay gain for different values of relay-destination distance.

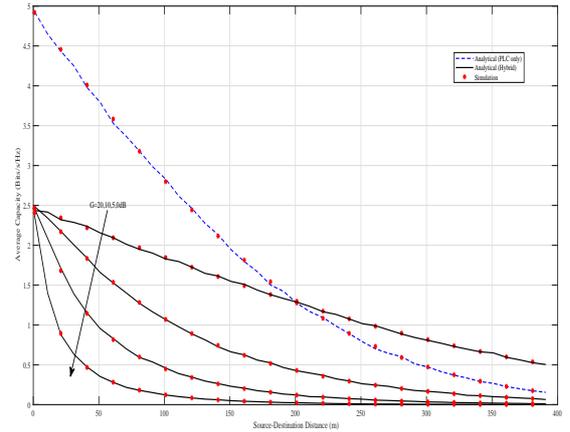

Fig. 5: Comparison between the hybrid system and PLC only system.

A comparison is carried out in Fig. 5 between the PLC system only and the hybrid system. Both $\sigma^2$ at PLC and hybrid system are equal to 0.1, $\sigma_r^2 = 0.01$ for the relay in the hybrid system, $d_1 = d_2$ both equal to the half of the total distance and $m = 2$. Different gain scenarios were considered and it was found that the PLC only system outperforms the hybrid system when the relay gain is varied from $0 - 10$dB regardless the source-destination distance whereas the hybrid system performance becomes higher than the performance of the former system when the relay gain is 20dB and distance between the source and destination is more than 200m.

## V. Conclusions

In this paper, we discussed the implementation of AF relaying in hybrid PLC/wireless communication systems for indoor applications. The system performance was analyzed in terms of the average capacity. The analytical expression of the capacity was validated by Monte Carlo simulation. The impact of some system parameters was investigated in this paper. The discussion showed that the performance of the proposed hybrid system can be enhanced by increasing the gain of the relay. It was found that the capacity of the system is affected by the relay-destination distance and it degrades as this distance increases. Moreover, the path exponent loss of the wireless link has an adverse impact on the system under consideration. The source-to-relay distance was also discussed and it was found that it does not have a considerable effect on the performance if the increase of the distance is relatively small. Although, the performance of the PLC only system outperforms that of the hybrid system particularly for low values of source-destination distance and relay gain, yet the use of the hybrid system is more attractive in indoor applications since it provides better mobility to end users.

## VI. Acknowledgment

This research has been jointly funded by OSL Rail Ltd, and the Faculty of Science and Engineering, Manchester Metropolitan University, UK.